\documentclass[aps,pra,twocolumn,superscriptaddress,nofootinbib]{revtex4-2}
\usepackage{graphicx}
\usepackage{amssymb}
\usepackage{amsmath}
\usepackage{bm}
\usepackage{xcolor}
\usepackage{verbatim}
\usepackage{footnote}
\usepackage{dsfont}
\usepackage{mathtools}
\usepackage{float}
\usepackage{xfrac}
\usepackage{dcolumn} 
\usepackage{braket}
\usepackage{soul}

\usepackage{hyperref}
\hypersetup{
     colorlinks=true,
     linkcolor=blue,
     filecolor=blue,
     citecolor=blue,  
     urlcolor=black,
     }
\usepackage[all]{hypcap}

\begin{document}
\title{Efficient ensemble randomization by tuning chaos in a nonlinear spin-1 system}

\author{Jongmin Kim}
\thanks{These authors contributed equally to this work.}
\affiliation{Department of Physics and Astronomy, Seoul National University, Seoul 08826, Korea}
\affiliation{NextQuantum, Seoul National University, Seoul 08826, Korea}

\author{Minsung Jeong}
\thanks{These authors contributed equally to this work.}
\affiliation{Department of Physics and Astronomy, Seoul National University, Seoul 08826, Korea}

\author{Jongyoon Han}
\affiliation{Department of Physics and Astronomy, Seoul National University, Seoul 08826, Korea}

\author{Y. Shin}
\email{yishin@snu.ac.kr}
\affiliation{Department of Physics and Astronomy, Seoul National University, Seoul 08826, Korea}
\affiliation{NextQuantum, Seoul National University, Seoul 08826, Korea}
\affiliation{Institute of Applied Physics, Seoul National University, Seoul 08826, Korea}

\date{\today}

\begin{abstract}
We present an efficient scheme to randomize a spin-state ensemble in a nonlinear spin‑1 system by tuning chaos with an external periodic drive. 
Without modulation, the system exhibits a mixed phase space featuring regular islands embedded in a chaotic sea, where global mixing is inhibited by energy conservation. 
Using numerical simulations, we demonstrate that modulation of a linear Zeeman field not only facilitates transport between different energy shells but also drives ensembles toward a Haar‑random distribution over spin states. 
Under optimized conditions, complete randomization at the level of the second moment is achieved on a timescale set by the inverse nonlinear interaction energy.
In the overdriven regime, randomization is unexpectedly suppressed at specific modulation amplitudes, accompanied by the formation of sticky regions in phase space. 
We attribute this behavior to the dynamical cancellation of the leading low-order harmonic component of the periodic drive. 
These results illustrate how time-periodic driving can be used to engineer chaotic systems and achieve controllable randomization in nonlinear spin systems.
\end{abstract}
\maketitle

\section{Introduction}

Understanding how deterministic many-body dynamics generate statistical behavior is a central problem in nonequilibrium physics. In both classical and quantum systems, highly random ensembles are crucial in studies of scrambling, thermalization, and state preparation~\cite{DAlessio16,Iyoda18,Brandao16}. In particular, ensembles approaching Haar-random statistics provide a benchmark for ergodicity and the loss of memory of initial conditions~\cite{PilatowskyCameo24}. However, generating truly random ensembles is experimentally demanding. In realistic platforms, one instead seeks dynamical protocols that efficiently erase preparation-dependent structures using a restricted set of controllable parameters~\cite{Choi23}. Identifying such protocols and quantifying their efficiency remains an important challenge.

A natural route toward randomization is to exploit chaotic dynamics~\cite{Eckmann85}. In systems exhibiting sensitivity to initial conditions, nearby trajectories diverge exponentially, leading to effective unpredictability over long timescales. However, chaos alone does not guarantee global randomization. While local instability is characterized by positive Lyapunov exponents, efficient randomization requires ergodic transport over the accessible phase space~\cite{Meiss92}. In particular, when motion is constrained to invariant manifolds such as energy shells, the system may remain confined to a limited region despite being locally chaotic. Thus, the distinction between chaos (local instability) and ergodicity (global exploration) is central to designing efficient randomization protocols.

In this study, we investigate a randomization scheme in the internal spin dynamics of a spin-1 Bose-Einstein condensate (BEC). The spin dynamics of a spin-1 condensate are governed by a nonlinear mean-field Hamiltonian, whose phase space generically exhibits a mixed structure, with regular and chaotic trajectories coexisting depending on the initial condition and control parameters~\cite{Rautenberg20}. 
When the Hamiltonian is time-independent, energy is conserved, and the system's trajectory remains within a single energy shell, thereby limiting phase-space exploration~\cite{Tabor89}. Recent work has shown that time-periodic modulation of control fields can lift this constraint and promote transport between energy shells, enabling ensembles of spin states to approach a Haar-random distribution~\cite{Kim24}. In spatially extended spinor BECs, the degree of global randomization of the corresponding single-mode spin dynamics was found to correlate with the development of spatially disordered spin textures and turbulence~\cite{Jung23}, which has recently been explored experimentally~\cite{Kim24,Hong23}. 


Motivated by this connection, here we investigate how complete randomization develops in this spin system within the single-mode approximation and how rapidly it can be achieved depending on the driving parameters.
Our goal is to identify efficient driving protocols that enable rapid and complete randomization of the spin state of a spin-1 spinor BEC.
To this end, we characterize the spin dynamics of the system as the driving amplitude is increased, explicitly distinguishing between chaoticity and randomization. Chaoticity is quantified through the largest Lyapunov exponent (LLE), which measures local exponential instability~\cite{Benettin80a,Skokos10}. Randomization is characterized using two complementary diagnostics: (i) the Shannon entropy for single trajectories, which probes phase-space coverage~\cite{Cincotta20}; and (ii) the trace distance between the ensemble second moment and its Haar value, which measures the approach to Haar-random statistics~\cite{Ho22,Cotler23}. These measures allow us to quantify the emergence of global mixing beyond local chaotic instability.

Using numerical simulations, we show that as the modulation amplitude increases, the volume of the chaotic sea in phase space gradually increases, and full mixing with global instability occurs when the modulation amplitude becomes comparable to the characteristic energy scale of the system.
Under optimized conditions, complete randomization is achieved on a timescale set by the inverse nonlinear interaction energy.
Furthermore, in an overdriven regime, we reveal that randomization is significantly suppressed at specific modulation amplitudes, accompanied by the emergence of sticky regions in phase space.  
We attribute this behavior to the dynamical cancellation of the low-order harmonic contribution from the periodic drive.

These results demonstrate that the chaotic dynamics of a spin-1 system can be systematically controlled by external periodic modulation, ranging from regimes of efficient randomization to regimes where chaos is partially suppressed. By establishing quantitative diagnostics and identifying optimal driving regimes, our work provides a framework for engineering random ensembles in nonlinear spin systems. More broadly, it highlights new opportunities for Floquet engineering of chaotic systems~\cite{Bukov15,Eckardt17} and offers insight into the interplay between chaos, ergodicity, and controlled scrambling in driven many-body dynamics~\cite{Mark24}.

\section{Model system}
\subsection{Periodically driven spin-1 BEC}
\label{PS}

We consider a spin-1 BEC in the regime where its order parameter can be written as a coherent wave function $\boldsymbol{\Psi} = \sqrt{n} e^{i\varphi}\boldsymbol{\zeta}$~\cite{Ho98,Kawaguchi12,StamperKurn13}. Here, $n$ is the number density, $\varphi$ is the global phase, and $\boldsymbol{\zeta}=\left(\zeta_{1},\zeta_{0},\zeta_{-1}\right)^{\mathrm{T}}$ is a normalized spinor describing the internal spin state. Each component can be written as  $\zeta_i = \sqrt{\rho_i}e^{i \theta_i }$, where $\rho_i$ and $\theta_i$ denote the population fraction and phase, respectively, of the $m_F=i$ component.

Within the single-mode approximation, which neglects spatial dynamics, we focus on the internal spin dynamics of the condensate \cite{Law98,Yi02}.
The Hamiltonian for the spin dynamics is given by 
\begin{equation}
    H_0(\boldsymbol{\zeta},\boldsymbol{\zeta}^\dagger) =  \hbar \Omega  f_x + q \boldsymbol{\zeta}^{\dagger} \text{F}_z^2 \boldsymbol{\zeta} +\frac{\varepsilon_s}{2} \vert \boldsymbol{f} \vert^2, \label{H0}
\end{equation}
where $\hbar$ is the reduced Planck constant, $\textbf{F}=(\text{F}_x,\text{F}_y,\text{F}_z)$ are the spin-1 operators, and $\boldsymbol{f} = \boldsymbol{\zeta}^\dagger \textbf{F} \boldsymbol{\zeta}$ is the spin vector, whose components correspond to magnetization in the $x$, $y$, and $z$ directions.
$H_0$ can be realized experimentally by applying a bias magnetic field along the $z$ direction together with a transverse rf magnetic field tuned to the Larmor resonance. In the corresponding rotating frame, the spin evolution is governed by Eq.~(\ref{H0})~\cite{Hong23}. Here, $\Omega$ denotes the Rabi frequency of the transverse rf field, $q$ is the quadratic Zeeman shift induced by the bias field, and $\varepsilon_s$ represents the spin-interaction energy arising from spin-dependent collisions.

\begin{figure}[t]
	\includegraphics[width=8.6cm]{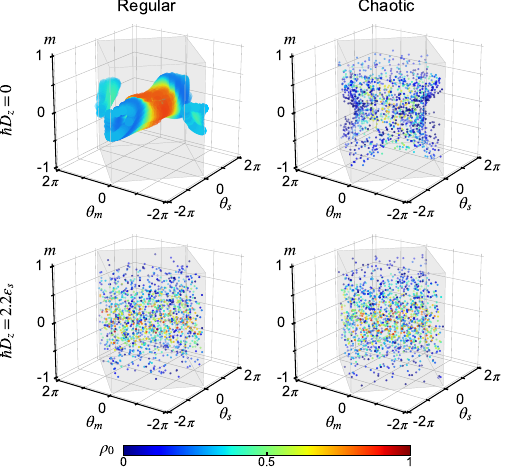}
	\caption{
    Phase-space portraits of representative regular and chaotic trajectories, initialized at $\boldsymbol{x}_R$ and $\boldsymbol{x}_C$, respectively. The plotted coordinates are $(m, \theta_s, \theta_m)$, while $\rho_0$ is encoded by the point color. Without modulation ($\hbar D_z/\varepsilon_s=0$, upper row), the motion remains confined to the constant-energy shell set by the initial condition. The chaotic trajectory densely explores a broad portion of the accessible manifold, whereas the regular trajectory is restricted to a lower-dimensional subset of it. In the driven case ($\hbar D_z/\varepsilon_s=2.2$, lower row), conservation of the static Hamiltonian is broken and both trajectories spread over a broader region of the diagram, indicating global phase-space mixing. For visualization, the trajectories are sampled at intervals of 67~ms $\approx3h/\varepsilon_s$, which is different from the sampling time used for the phase-space coverage calculation. 
     }
	\label{FIG1}
\end{figure}

The nonlinear form of $H_0$ yields complex dynamical behavior. In particular, when the energy scales associated with $q$, $\varepsilon_s$, and $\hbar \Omega$ are comparable, the system is known to exhibit chaotic dynamics~\cite{Rautenberg20}. 
Throughout this work, we use the set of parameters $q/h=\varepsilon_s/h=45~\mathrm{Hz}$ and $\Omega/2\pi=22.5~\mathrm{Hz}$, motivated by recent experiments~\cite{Hong23,Kim24}.
Since $H_0$ is time-independent, the energy of the system is conserved during evolution.
As a result, the system's trajectory in phase space remains confined to a single constant-energy shell, which prevents full exploration of the phase space and therefore limits ensemble randomization.

In this work, we investigate the spin dynamics of the system under a time-periodic drive with
\begin{align}
    &H(\boldsymbol{\zeta},\boldsymbol{\zeta}^\dagger ;t) = H_0 + H_F (t), \label{H} \\
    &H_F(t) = \hbar D_z \sin(\omega_m t ) f_z, \label{HF}
\end{align}
which can be implemented by modulating the bias field along $z$ or applying a frequency modulation to the rf field \cite{Hong23}.
The driving term $H_F(t)$ acts as a controllable periodic perturbation that modifies the chaoticity of the dynamics and enables transport between different energy shells, possibly promoting ensemble randomization in phase space.

\subsection{Phase-space representation of spin states}

To analyze the dynamics of the system, we introduce a classical phase-space representation of the spin state. 
The spin state satisfies the normalization condition $\boldsymbol{\zeta}^{\dagger}\boldsymbol{\zeta}=1$ and possesses a gauge redundancy $\boldsymbol{\zeta}\sim e^{i\chi}\boldsymbol{\zeta}$.
Consequently, the physical state space is the complex projective plane $\mathbb{C}\mathrm{P}^{2}$, a four-dimensional real manifold \cite{Bengtsson17}.
Accordingly, we parameterize the state using four variables $(\rho_{0},m,\theta_{s},\theta_{m})$ defined as~\cite{Rautenberg20}
\begin{eqnarray}
    m = \rho_1 - \rho_{-1},\\
    \theta_s = \theta_1 +\theta_{-1}  - 2\theta_0, \\
    \theta_m = \theta_1 - \theta_{-1}.
    \label{StateVariable}
\end{eqnarray}
Here, $m$ is the longitudinal magnetization, and $\theta_s$ and $\theta_m$ are relative phase variables.
From $\rho_1+\rho_0+\rho_{-1}=1$ and $\rho_i\geq 0$, $\rho_0 \in [0, 1]$ and $m \in [-1+\rho_0,\,1-\rho_0]$, and the $2\pi$ periodicity of $\theta_i$ sets a periodic rhombic domain for the angular sector $(\theta_s,\theta_m)$, as illustrated in Fig.~\hyperref[FIG1]{1}.

In this representation, the dynamics of the spinor BEC system is described as a trajectory $\boldsymbol{x}(t)$ in the phase space with coordinates $(\rho_0,m, \theta_s, \theta_m)$. 
From the time evolution of $\boldsymbol{\zeta}$, $\frac{\partial \boldsymbol{\zeta}}{\partial t} = -\frac{i}{\hbar}\frac{\partial H(\boldsymbol{\zeta},\boldsymbol{\zeta}^\dagger ;t)}{\partial \boldsymbol{\zeta}^\dagger}$,
we derive the corresponding equations of motion,
\begin{equation}
\begin{aligned}
    \frac{\partial \rho_0 }{\partial t} =  -\frac{2}{\hbar} \frac{\partial H}{\partial \theta_s }&,~~\; 
    \frac{\partial \theta_s  }{\partial t} =  \frac{2}{\hbar} \frac{\partial H}{\partial \rho_0 },\\
    \frac{\partial m }{\partial t} =  \frac{2}{\hbar} \frac{\partial H}{\partial \theta_m  }&,~~\; 
    \frac{\partial \theta_m  }{\partial t} =  -\frac{2}{\hbar} \frac{\partial H}{\partial m },
\label{EoM_PS}
\end{aligned}
\end{equation} 
where the Hamiltonian takes the form
\begin{align} 
    H_0 &= \hbar \Omega \sqrt{\rho_0} \bigg[ \sqrt{1-\rho_0+m} \cos{\left(\frac{\theta_s + \theta_m }{2}\right)} \nonumber \\ 
    &\quad + \sqrt{1-\rho_0-m} \cos{\left(\frac{\theta_s - \theta_m }{2}\right)} \bigg] \nonumber \\
    &\quad + q \left(1-\rho_0 \right) + \varepsilon_s \rho_0 (1-\rho_0 ) \nonumber \\
    &\quad + \varepsilon_s \left( \rho_0 \sqrt{(1-\rho_0 )^2 -m^2} \cos{\theta_s } + \frac{m^2 }{2} \right)
    \label{Hamiltonian_PS} \\
    H_F&(t)
    = \hbar D_z \sin(\omega_m t)\, m .
    \label{HF_PS_z}
\end{align}

In the following sections, we analyze how the periodic drive modifies the phase-space structure and enables efficient randomization of spin-state ensembles.

\section{Results and discussion}

\subsection{Constrained chaotic trajectory}

We first consider the unmodulated system with $H(t)=H_0$. In this case, the dynamics sensitively depends on the initial condition: trajectories can be either regular, exhibiting quasiperiodic motion, or chaotic, characterized by strong sensitivity to initial conditions. 
Figure~\hyperref[FIG1]{1} shows three-dimensional phase-space portraits for two representative trajectories with initial conditions $\boldsymbol{x}_R=(0.51,0.25,0.85\pi,0.14\pi)$ and $\boldsymbol{x}_C=(0.70,0.28,0\pi,0\pi)$, corresponding to regular and chaotic motion, respectively. The regular trajectory remains confined to a lower-dimensional subset of phase space, whereas the chaotic trajectory densely explores a much broader portion of the accessible manifold.
Despite this contrast in dynamical behavior, both trajectories remain restricted to their respective constant-energy shells owing to energy conservation.

\begin{figure}[t]
	\includegraphics[width=8.6cm]{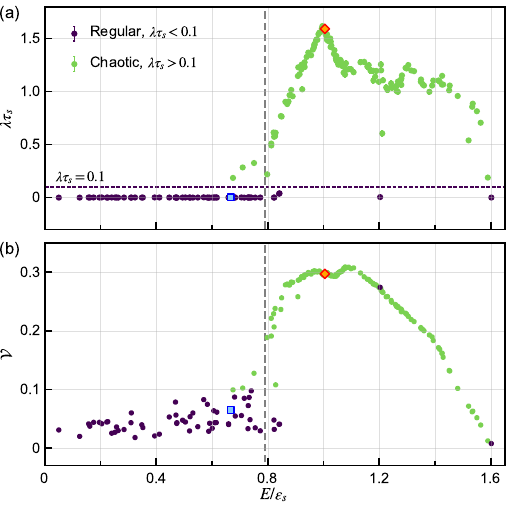}
	\caption{Mixed phase space without modulation. (a) Largest Lyapunov exponent (LLE), $\lambda$, and (b) phase-space coverage fraction, $\mathcal{V}$, for single trajectories as functions of the spin-state energy $E$. A total of 200 initial conditions were independently sampled from the Haar ensemble over pure spin-1 states.
    Here, $\tau_s=h/\varepsilon_s$ is the characteristic spin-interaction time.
    Regular (navy circles) and chaotic (green circles) trajectories are largely separated in energy, with an empirical crossover at $E/\varepsilon_s \approx 0.8$ (dashed line).
    The data points for the representative regular and chaotic initial conditions, $\boldsymbol{x}_R$ and $\boldsymbol{x}_C$ in Fig. 1, are indicated by blue and red markers, respectively.
    }	\label{FIG2}
\end{figure}

To quantitatively characterize the chaotic dynamics, we employ two complementary metrics that probe distinct aspects of a single trajectory: the LLE and the Shannon entropy.
The LLE, denoted by $\lambda$, quantifies the local dynamical instability by measuring the exponential growth of an infinitesimal displacement between nearby trajectories~\cite{Benettin80a},
\begin{equation}
    \lambda = \lim_{t\rightarrow\infty}\lim_{d(0) \rightarrow0} \frac{1}{t}\ln \frac{d(t)}{d(0)},
    \label{LyaDef}
\end{equation}
where $d(t)$ is the phase-space distance at time $t$. In this work, we define the phase-space distance as $d(t)\equiv\sqrt{(\Delta\rho_0)^2+(\Delta m)^2+(\Delta\theta_+/\pi)^2+(\Delta\theta_-/\pi)^2}$, 
where $\Delta\theta_{\pm} \in [-\pi,\pi)$ denotes the shortest angular displacement, taking into account the $2\pi$ periodicity of $\theta_{\pm} \equiv \theta_{\pm1}-\theta_0$.
A positive $\lambda$ indicates chaotic dynamics, whereas $\lambda\simeq 0$ corresponds to regular motion. In the evaluation of $\lambda$, we use the standard rescaling method~\cite{Skokos10}, the details of which are described in Appendix~A.

Although the LLE captures local instability, it does not quantify how extensively a trajectory explores phase space. To address this, we introduce the Shannon entropy as a complementary measure of phase-space coverage and ergodicity~\cite{Cincotta20}. 
We estimate the entropy $S$ of a single trajectory from its coarse-grained occupation over uniformly discretized phase-space cells. Specifically, we sample $N_s$ points along the trajectory $\{\boldsymbol{x}(t_k)\}_{k=1}^{N_s}$ with $t_k=kT_{\textrm{samp}}$, and compute
\begin{equation}
    S = - \sum_{i=1}^{N_{\mathrm{bin}}} p_i \log p_i
    = - \sum_{i=1}^{N_{\mathrm{bin}}} \frac{n_i}{N_s} \log \frac{n_i}{N_s},
    \label{ShannonEntropy}
\end{equation}
where $n_i$ is the number of samples in the $i$-th cell and $p_i=n_i/N_s$. Throughout this work, we use $T_{\textrm{samp}}=1~\mathrm{ms}=0.045\tau_s$ and total sampling time $t=100~\mathrm{s}=4500\tau_s$ ($N_s=100{,}000$), with $N_{\mathrm{bin}}=4096(=8^4)$ chosen to balance spatial resolution and statistical noise (see Appendix~A). Here, $\tau_s = h/\varepsilon_s\approx 22$~ms is the characteristic time set by the intrinsic interaction energy scale of the system.

Because the absolute value of $S$ depends on the coarse-graining scheme, we instead consider the entropy deficit relative to the Haar-random distribution,
\begin{equation}
    \Delta S \equiv S_{\mathrm{Haar}}- S,
    \label{EntropyDeficit}
\end{equation}
where $S_{\mathrm{Haar}}$ is the entropy obtained from $N_s$ samples drawn from the unitarily invariant Haar measure over pure spin-1 states~\cite{Ho22,Zyczkowski01}. By construction, $\Delta S \ge 0$, and $\Delta S\simeq 0$ indicates maximal coverage at the chosen resolution. We further define an effective phase-space coverage fraction as $\mathcal{V}=\exp(-\Delta S)\leq 1$.

Having established these diagnostics, we now map the global phase-space structure of the system. We uniformly sample 200 initial states from the Haar ensemble and compute $\lambda$ and $\mathcal{V}$ for each trajectory. Figure~\hyperref[FIG2]{2} shows the resulting distributions as functions of the spin-state energy $E$, calculated using Eq.~(\ref{Hamiltonian_PS}).
Both quantities exhibit a clear crossover at $E_{\mathrm{c}}/\varepsilon_s \approx 0.8$. For $E<E_{\mathrm{c}}$, the LLE remains near zero while $\mathcal{V}$ scatters in a low region of $\mathcal{V}<0.2$, consistent with trajectories confined to regular islands. In contrast, for $E>E_{\mathrm{c}}$, the LLE becomes positive and $\mathcal{V}$ collapses to a nearly single-valued curve at fixed energy, indicating that chaotic trajectories explore their accessible energy shells much more uniformly. The maximum LLE reaches $\lambda_0 = 1.62/\tau_s$ at $E/ \varepsilon_s \approx 1.0$.

\begin{figure}[t]
	\includegraphics[width=8.6cm]{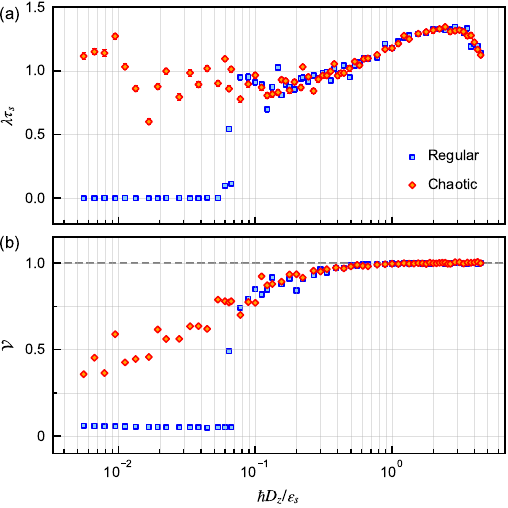}
	\caption{
    Evolution of trajectories with modulation. (a) $\lambda$ and (b) $\mathcal{V}$ for the trajectories initialized at $\boldsymbol{x}_R$ (blue) and $\boldsymbol{x}_C$ (red) as functions of the modulation amplitude $D_z$. The trajectory for $\boldsymbol{x}_R$ becomes chaotic with $\hbar D_z/\varepsilon_s \gtrsim 0.07$. The horizontal dashed line in (b) indicates $\mathcal{V}=1$, full coverage of the phase space.}
	\label{FIG3}
    \vspace{-4pt}
\end{figure}

\begin{figure}[t]
	\includegraphics[width=8.6cm]{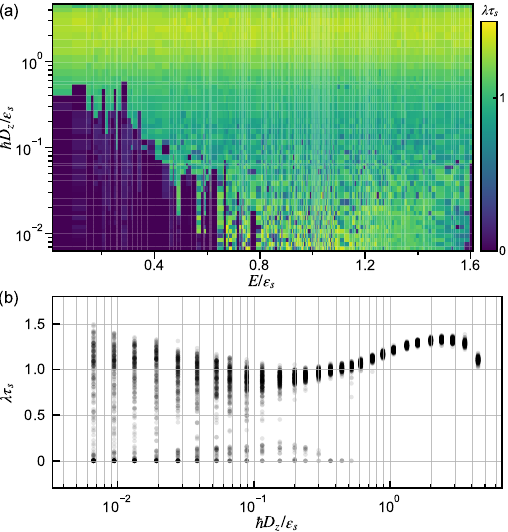}
	\caption{Global phase-space mixing induced by modulation. (a) Color map of the LLE, $\lambda$, on the $E-D_z$ plane. The map was constructed from the trajectories with the same 200 Haar-sampled initial conditions used in Fig.~\ref{FIG2}. Here, $E$ is the initial state energy.
    (b) LLE distribution as a function of $D_z$. 
    As the modulation amplitude increases, the regular portion is progressively suppressed. At $\hbar D_z /\varepsilon_s \gtrsim 0.6$, the LLE becomes positive for all sampled initial conditions, and its spread is strongly reduced, indicating that the local instability becomes uniform across the sampled phase space.}
	\label{Fig4}
\end{figure}

\subsection{Global mixing via modulation}
We now introduce a periodic modulation described by $H_F(t) = \hbar D_z \sin(\omega_m t)\, f_z.$
Unless otherwise stated, we fix the modulation frequency to $\omega_m/2\pi = 60~\mathrm{Hz}$ and investigate how the chaoticity of the system evolves as the modulation amplitude $D_z$ increases.

We first examine the responses of representative trajectories. Figure~\hyperref[FIG3]{3} shows the evolution of the LLE $\lambda$ and phase-space coverage fraction $\mathcal{V}$ for trajectories initialized at $\boldsymbol{x}_R$ (regular) and $\boldsymbol{x}_C$ (chaotic). For the chaotic initial state $\boldsymbol{x}_C$, $\lambda$ initially decreases slightly with increasing $D_z$, reflecting a reorganization of local stability as neighboring energy shells begin to mix, while $\mathcal{V}$ increases. In contrast, for the regular initial state $\boldsymbol{x}_R$, both $\lambda$ and $\mathcal{V}$ remain nearly constant at small modulation amplitudes and then rise sharply at $\hbar D_z \approx 0.07\varepsilon_s$, reaching values close to those of $\boldsymbol{x}_C$, signaling escape from the regular island and the onset of chaotic dynamics. 

At larger amplitudes, $\hbar D_z \gtrsim 0.2\varepsilon_s$, $\lambda$ increases further and reaches a maximum $\lambda \approx 1.3/\tau_s$ near $\hbar D_z \approx 2.2\varepsilon_s$, while $\mathcal{V}$ approaches unity for both initial conditions. This demonstrates that the modulation not only enables inter-shell transport but also converts regular motion into chaotic dynamics, suggesting a substantial expansion of chaotic regions in phase space.

To establish the global nature of this transition, we extend the analysis to an ensemble of initial states. Using the same 200 Haar-sampled initial conditions as in Fig.~\hyperref[FIG2]{2}, we compute the LLEs for various $D_z$, and the result is shown in Fig.~\hyperref[Fig4]{4(a)} as a function of $D_z$ and $E$. As $D_z$ increases, the regular part of the ensemble is progressively reduced with decreasing crossover energy $E_c$, which separates the regular and chaotic dynamics. Once the modulation amplitude exceeds $\hbar D_z \approx 0.6\varepsilon_s$, all sampled LLEs become positive, indicating that chaotic dynamics fully dominates the system.

Figure~\hyperref[Fig4]{4(b)} shows a plot of the same data to illustrate how the LLE distribution evolves with increasing $D_z$. While the trajectories of many initial states that are regular in the unmodulated case acquire positive Lyapunov exponents, some initial states with chaotic trajectories show a slight reduction in $\lambda$ at small $D_z$, consistent with the behavior observed for $\boldsymbol{x}_C$. When regular trajectories are fully eliminated for  $\hbar D_z \gtrsim 0.6\varepsilon_s$, the LLE becomes nearly uniform across the phase space. 
This homogenization is consistent with the formation of a broadly connected chaotic sea and indicates robust global mixing across the sampled phase space.
We note that the maximum LLE in the driven system, $\lambda \approx 1.3/\tau_s$, remains comparable to the peak value in the undriven case, $\lambda_0 = 1.62/\tau_s$.


\begin{figure*}[t]
	\includegraphics[width=14.8cm]{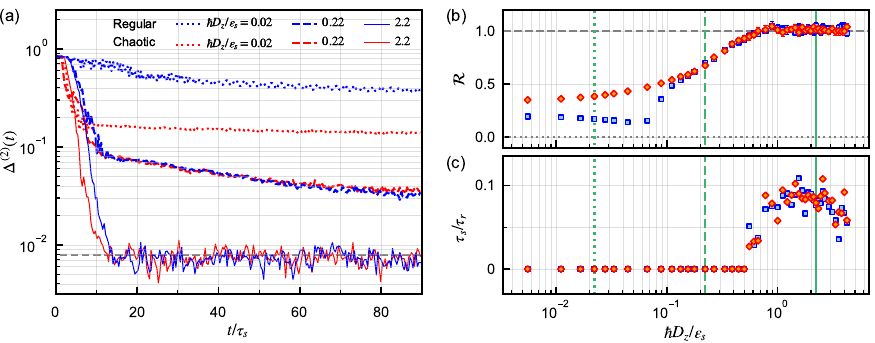}
	\caption{Ensemble randomization in the periodically driven spin system: weak modulation regime. (a) Time evolution of the trace distance $\Delta^{(2)}(t)$ for ensembles $\mathcal{E}_R$ and $\mathcal{E}_C$, initialized near the representative spin states, $\boldsymbol{x}_R$ (regular) and $\boldsymbol{x}_C$ (chaotic), respectively, for different modulation amplitudes $D_z$. The dashed horizontal line indicates the finite-size floor $\Delta_m^{(2)} = 1/\sqrt{N_{\mathrm{ens}}}=0.0078$, where $N_{\mathrm{ens}}=128^2$ is the number of spin states in the ensemble. (b) Ensemble randomness $\mathcal{R}$ and (c) randomization time $\tau_r$ as functions of $D_z$ for $\mathcal{E}_R$ (blue) and $\mathcal{E}_C$ (red). $\tau_r$ was determined as the first time at which the trace distance reaches $\Delta_m^{(2)}$ (see the text). The vertical lines indicate the values of $D_z$ in (a). 
    }
	\label{Fig5}
\end{figure*}

\subsection{Ensemble randomization}

We now investigate the ensemble randomization process under periodic driving. Although the spin state of a BEC is represented as a point in classical phase space, experimentally prepared coherent spin states inevitably possess finite phase uncertainty due to both technical noise and intrinsic quantum fluctuations~\cite{Yi03,Leslie09}. Within a semiclassical description, this uncertainty can be modeled as an ensemble with a finite phase-space volume, which allows us to define and quantify the timescale for full randomization.

To this end, we consider the time evolution of a spin-state ensemble centered on a given initial state. The ensemble  consists of $N_{\mathrm{ens}}=128^2$ spin states with an initial spread $d_i \sim 5\times10^{-3}$ in the phase-space metric, which is generated by adding Gaussian noise~\cite{Kim24}. The spread corresponds to the quantum noise level for a condensate of $\approx 10^4$ atoms.
To quantify the degree of randomization, we employ the trace distance~\cite{Ho22,Cotler23},
\begin{eqnarray}
    \Delta^{(2)} \equiv \frac{1}{2}\left\| \rho_{\mathcal{E}}^{(2)}-\rho_{\text{Haar}}^{(2)} \right\|_{1},
    \label{TD}
\end{eqnarray}
where $\rho_{\mathcal{E}}^{(2)}=\frac{1}{N_{\mathrm{ens}}}\sum_{\boldsymbol{\zeta}\sim\mathcal{E}} \left(\boldsymbol{\zeta}\boldsymbol{\zeta}^{\dagger}\right)^{\otimes 2}$ is the second moment of the ensemble $\mathcal{E}$, and $\rho_{\text{Haar}}^{(2)}$ is that of the Haar ensemble, i.e., the unitarily invariant maximally randomized distribution over pure spin-1 states. Here, $\|\cdot\|_1$ denotes the trace norm, and $0\le \Delta^{(2)}\le1$ quantifies the deviation from the Haar ensemble, with smaller values indicating a higher degree of randomization. 
Owing to finite sampling, the trace distance cannot vanish exactly. Guided by the finite-sampling scaling of projected ensembles~\cite{Cotler23}, we use \(\Delta_{m}^{(2)} = 1/\sqrt{N_{\mathrm{ens}}}=0.0078\) as the finite-size reference.
By tracking the time evolution of $\Delta^{(2)}(t)$, we characterize both the extent and rate of ensemble randomization.

Figure~\hyperref[Fig5]{5(a)} shows $\Delta^{(2)}(t)$ for ensembles $\mathcal{E}_R$ and $\mathcal{E}_C$ centered at $\boldsymbol{x}_R$ and $\boldsymbol{x}_C$, respectively, for various modulation amplitudes. In the weak-modulation limit ($\hbar D_z/\varepsilon_s=0.02$), the ensemble initialized in the chaotic region exhibits a faster initial decay due to its intrinsic instability. However, in both cases, $\Delta^{(2)}(t)$ saturates at values significantly above $\Delta_m^{(2)}$, indicating incomplete randomization under energy-shell confinement. 
With increasing modulation amplitude, randomization is strongly enhanced for both initial conditions. Notably, for sufficiently large amplitude ($\hbar D_z/\varepsilon_s=2.2$), both ensembles approach the finite-size floor, $\Delta^{(2)}\simeq \Delta_m^{(2)}$, demonstrating full randomization.

To quantify the efficiency of this process in terms of both the extent of randomization and the time required to achieve it, we introduce two metrics: ensemble randomness $\mathcal{R}$ and the randomization time $\tau_r$. The randomness is defined by
\begin{equation}
\mathcal{R} \equiv \frac{\log\left[\Delta^{(2)}_{p}/\Delta^{(2)}(t_f)\right]}{\log\left[\Delta^{(2)}_{p}/\Delta^{(2)}_{m}\right]},
\label{randomness}
\end{equation}
where $\Delta_{p}^{(2)}=\frac{5}{6}$ is the trace distance between any pure spin-1 state and the Haar ensemble.\footnote{For an ensemble consisting of a single pure state, one can choose a basis such that $\rho_\varepsilon^{(2)}=\operatorname{diag}(1,0_{8\times8})$ and $\rho_{\mathrm{Haar}}^{(2)}=\frac{1}{6}\operatorname{diag}(I_6,0_{3\times3})$~\cite{Cotler23}, yielding $\Delta_{p}^{(2)}=\frac{1}{2}\left(\left|\frac{5}{6}\right|+5\left|-\frac{1}{6}\right|\right)=\frac{5}{6}$ from Eq.~(\ref{TD}).} With this definition, $\mathcal{R}=0$ for an ensemble localized at a single pure state and $\mathcal{R}=1$ when the ensemble reaches the finite-size floor. The randomization time $\tau_r$ is defined as the first time at which the trace distance reaches the finite-size floor $\Delta^{(2)}_m$. Here, $t_f=90\tau_s$ is chosen to be sufficiently long to determine the steady state. If $\Delta^{(2)}(t)>\Delta^{(2)}_{m}$ throughout the interval $0\leq t\leq t_f$, we set $\tau_r=\infty$.


Figures~\hyperref[Fig5]{5(b)} and \hyperref[Fig5]{5(c)} show $\mathcal{R}$ and $\tau_r$, respectively, as functions of $D_z$. As $D_z$ increases, the distinction between ensembles initialized at $\boldsymbol{x}_R$ and $\boldsymbol{x}_C$ progressively diminishes: their $\mathcal{R}$ values collapse for $\hbar D_z\gtrsim0.2\varepsilon_s$, consistent with the merging of regular regions into the chaotic sea observed in the single-trajectory analysis. 
Full randomization is achieved for $\hbar D_z\gtrsim0.6\varepsilon_s$, where both ensembles approach $\mathcal{R} \simeq 1$. Correspondingly, the randomization time $\tau_r$ becomes finite and reaches a minimum $\tau_r\approx 12\,\tau_s$ at $\hbar D_z\approx 2.2\varepsilon_s$, indicating optimal randomization efficiency.
This timescale can be understood in terms of the Lyapunov instability of the system. Starting from an initial spread $d_i \sim 5\times10^{-3}$ and requiring expansion to a phase-space scale $d_f \sim 1$, the characteristic divergence time is estimated as $\tau_r \sim \frac{1}{\lambda}\ln\!\left(\frac{d_f}{d_i}\right) \sim 4\,\tau_s$,
comparable to the numerically observed optimal value of $\tau_r$. 

These results demonstrate that weak periodic driving provides an efficient and controllable route to global randomization. Once the modulation amplitude reaches $\hbar D_z\sim 2\varepsilon_s$, the system forms a fully connected chaotic sea, and ensembles rapidly converge to the Haar-random distribution on a timescale set by the intrinsic chaoticity of the nonlinear system. This timescale remains of order $\tau_s$, the intrinsic timescale set by the nonlinear spin interaction, rather than becoming parametrically longer.

\begin{figure*}[t]
	\includegraphics[width=14.8cm]{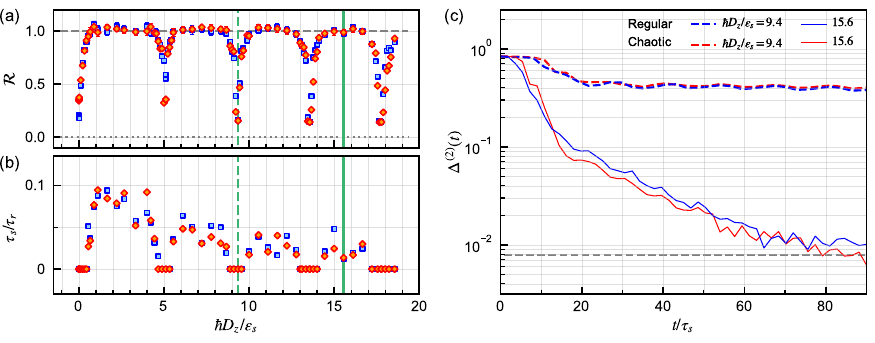}
	\caption{Suppression of ensemble randomization in the overdriven regime. (a) Ensemble randomness $\mathcal{R}$ and (b) randomization time $\tau_r$ over a broader range of $D_z$. 
    Sharp dips in $\mathcal{R}$ and divergences of $\tau_r$ mark modulation amplitudes where full randomization is not reached within the observation window. (c) Time evolution of the trace distance $\Delta^{(2)}(t)$ for two different modulation amplitudes $D_z$ (vertical lines in (a) and (b)), for the ensembles $\mathcal{E}_R$ (blue) and $\mathcal{E}_C$ (red) in Fig.~\ref{Fig5}. The horizontal dashed line indicates the finite-size floor $\Delta^{(2)}_m$. }
	\label{Fig6}
\end{figure*}

\subsection{Overdriven regime}

Finally, we examine the system’s response in the overdriven regime. As shown in Fig.~\hyperref[Fig5]{5(c)}, the randomization process is most efficient near $\hbar D_z \approx 2.2\varepsilon_s$. Beyond this optimal point, further increases in $D_z$ lead to a progressive slowdown of the dynamics, reflected in a longer randomization time $\tau_r$.
Figure~\hyperref[Fig6]{6} presents the ensemble randomization data over a broader range of modulation amplitudes, extending up to $\hbar D_z \sim 20\varepsilon_s$. In this regime, both $\mathcal{R}$ and $\tau_r$ exhibit pronounced nonmonotonic behavior [Figs.~\hyperref[Fig6]{6(a)--(b)}]. While the overall trend indicates reduced randomization efficiency at large $D_z$, we observe a series of points where $\mathcal{R}$ remains small and $\tau_r$ fails to reach a finite value within the observation window. A representative case at $\hbar D_z = 9.4\varepsilon_s$ is shown in Fig.~\hyperref[Fig6]{6(c)}, where the ensemble remains far from a fully randomized distribution regardless of the initial conditions. Notably, this suppression is not isolated but recurs systematically with increasing $D_z$, forming a sequence of sharp dips superimposed on a slowly varying envelope.

\begin{figure*}[t]	
\includegraphics[width=17.8cm]{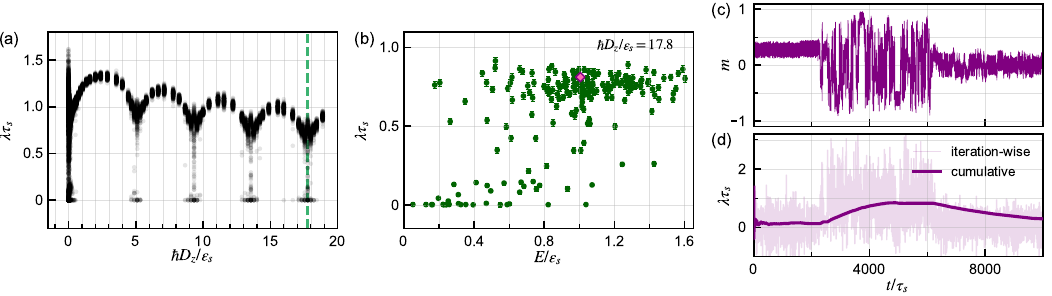}
	\caption{
    LLE distribution and intermittent dynamics in the overdriven regime. (a) LLE distribution of 200 Haar-sampled initial conditions as a function of $D_z$. The overall envelope decreases with increasing modulation amplitude, while sharp dips mark amplitudes at which the randomization is strongly suppressed. The green line in (a) indicates the representative dip analyzed in panels (b)--(d). (b) LLE as a function of initial energy $E$ at $\hbar D_z/\varepsilon_s = 17.8$. The highlighted point denotes the trajectory used in panels (c)--(d). (c) Time trace of the magnetization $m$ for the highlighted trajectory, showing intermittent trapping in and escape from sticky regions in phase space. (d) Corresponding iteration-wise (light line) and cumulative (bold line) estimate of the LLE. During regular-like intervals, the iteration-wise exponents fluctuate around zero, causing the cumulative average to decrease gradually.}
	\label{Fig7}
\end{figure*}

This anomalous behavior is corroborated by single-trajectory diagnostics. Figure~\hyperref[Fig7]{7}(a) shows the LLE distribution computed from 200 initial states. As $D_z$ increases, the overall magnitude of the LLE decreases, with notable dips occurring at the same modulation amplitudes identified in $\mathcal{R}$. At these points, the spread of the LLE values increases significantly, indicating that the dynamics is no longer uniformly chaotic across phase space.
Further insight is obtained from Fig.~\hyperref[Fig7]{7(b)}, which plots the LLE as a function of the initial-state energy $E$ at $\hbar D_z = 17.8\varepsilon_s$ (fourth dip). In contrast to the undriven case, where regular and chaotic regions are clearly separated by an energy-dependent crossover, the phase space here lacks a simple energy-based organization. Instead, chaotic and weakly unstable trajectories are interspersed throughout the phase space without a clear global structure.

At the level of individual trajectories, the dynamics often exhibits marked intermittency. As illustrated in Figs.~\hyperref[Fig7]{7(c)} and \hyperref[Fig7]{7(d)}, trajectories can remain trapped in regular-like motion with finite magnetization for extended periods, followed by transitions to chaotic dynamics with finite LLE, and subsequently return to regular-like behavior. This intermittency makes the LLE sensitive to the observation window, reflecting the alternation between trapped and chaotic intervals. 
The emergence of this sticky dynamics implies that at specific modulation amplitudes, trajectories undergo long-lived trapping near localized phase-space structures. Such stickiness is associated with long dwell times and slow transport in mixed Hamiltonian dynamics, thereby suppressing global exploration and hindering full randomization~\cite{Altmann06,Zou07,Lozej20,Altmann13}.

To elucidate the origin of this suppression, we analyze the system in a comoving frame defined by the unitary transformation $U(t) = \exp\!\left[i \tilde{D}_z \cos(\omega_m t) \, \text{F}_z \right]$, 
where $\tilde{D}_z = D_z/\omega_m$. The transformed  mean-field Hamiltonian reads
$H_I(t)= q \boldsymbol{\zeta}^{\dagger} \text{F}_z^2 \boldsymbol{\zeta}
    + \frac{\varepsilon_s}{2} |\boldsymbol{f}|^2 + \tilde{H}_F(t)$ with 
\begin{align}
    \tilde{H}_F(t) &=
    \hbar\Omega\bigg(\!
    \cos\!\left[\tilde{D}_z\cos(\omega_m t)\right] f_x + \sin\!\left[\tilde{D}_z\cos(\omega_m t)\right] f_y
    \bigg) \nonumber\\
     &=  \hbar\Omega\, J_0(\tilde{D}_z) f_x \nonumber\\
    & + 2\hbar\Omega\sum_{n=1}^{\infty}(-1)^n\bigg[
    J_{2n}(\tilde{D}_z)\cos\!\big(2n\cdot \omega_m t\big)\, f_x \nonumber\\
    &\quad
    - J_{2n-1}(\tilde{D}_z)\cos\!\big((2n-1)\cdot \omega_m t\big)\, f_y
    \bigg],
    \label{InteractionHamil}
\end{align}
where $J_n$ denotes the $n$-th order Bessel function of the first kind. In this frame, the longitudinal modulation is mapped onto transverse oscillations in the $xy$ plane, with a renormalized static field $\Omega_{\mathrm{eff}}=\Omega J_0(\tilde{D}_z)$ and additional higher-harmonic components.

For $\tilde{D}_z \gg 1$, the time-periodic term $\tilde{H}_F(t)$ can be regarded as a rapidly rotating transverse field. When its rotation frequency exceeds the characteristic energy scale of the system, its influence on the spin dynamics is effectively suppressed through time averaging~\cite{Brinkmann16,Rahav03}. This provides a natural explanation for the gradual decrease in the LLE at large $\tilde{D}_z$. However, this time-averaging effect does not account for the abrupt breakdown of randomization observed at specific values of $\tilde{D}_z$.

We find that the breakdown amplitudes coincide with the zeros of the first-order Bessel function, $J_1(\tilde{D}_z)=0$. This correspondence suggests that the leading low-order oscillating component plays a central role in sustaining global phase-space transport. Notably, even when the static component vanishes ($\Omega_{\mathrm{eff}}=0$), complete randomization can still be achieved if the oscillating channels remain active, underscoring the essential role of low-order harmonics in driving global mixing. 

When this primary channel is suppressed, transport across energy shells becomes inefficient, even in the presence of higher-order harmonics. Consistent with this interpretation, we find that the optimal randomization efficiency is largely insensitive to the modulation frequency within an intermediate range, but is noticeably suppressed for $\omega_m/2\pi \gtrsim 90~\mathrm{Hz}$ (see Appendix B).
The amplitude-selective suppression observed here is reminiscent of Bessel-function renormalization in strongly driven systems, where specific dynamical channels are selectively weakened or eliminated near the zeros of Bessel functions~\cite{Dunlap86,Grifoni98,Lignier07}.

The emergence of sticky dynamics near these suppression points can be attributed to the dominance of higher-harmonic components. Although these high-frequency terms are inefficient at mediating global transport, they can still induce local irregular motion, leading to the formation of long-lived metastable structures in phase space. As a result, trajectories become intermittently trapped, giving rise to sticky dynamics and incomplete randomization, as observed.

Finally, we note that the dip structure exhibits a clear dependence on the modulation frequency and direction (Appendices B and C): both its visibility and its correspondence with the zeros of $J_1$ are most pronounced near $\omega_m/2\pi = 60~\mathrm{Hz}$ and for $z$-directional modulation. This dependence suggests an additional layer of dynamical control, the underlying mechanism of which warrants further investigation.

\section{Summary and outlook}

We have investigated controllable randomization in the internal spin dynamics of a spin-1 spinor BEC, focusing on the evolution of trajectory ensembles toward Haar-random statistics on the spin-state manifold. We showed that a sinusoidal modulation of control fields efficiently promotes phase-space mixing by enabling transport across energy shells. By tuning the modulation amplitude, the system can be continuously driven from shell-confined dynamics to a regime of global randomization.
We also found that in the overdriven regime, trajectories exhibit long-lived sticky dynamics associated with localized phase-space structure, which hinders global transport and significantly reduces randomization efficiency.
By introducing quantitative diagnostics and identifying optimal driving conditions, our work establishes a systematic framework for engineering and controlling randomization in nonlinear spin systems. 

An important direction for future work is to extend this study beyond the mean-field (coherent-state) description into the regime of quantum chaos. In this fully quantum setting, one expects the emergence of genuine many-body scrambling, characterized by the growth of entanglement and the approach to unitary designs~\cite{Ho22}. It will be particularly interesting to investigate how the classical mechanisms identified here, such as shell mixing and the emergence of sticky behavior, manifest in quantum observables, and whether optimal driving conditions can be exploited to realize fast and controllable generation of quantum-random states in experimentally accessible cold-atom platforms. Another important extension is to investigate how phase-space randomization within the single-mode dynamics is connected to the emergence of spatially resolved random spin textures and turbulence~\cite{Kim24,Jung23,Hong23}.

\begin{acknowledgments}
We thank Junghoon Lee and Donggyu Lee for insightful discussion. 
This work was supported by the National Research Foundation of Korea (Grants No. RS-2023-NR077280, No. RS-2023-NR119928, and No. RS-2024-00413957).

\end{acknowledgments}

\section*{Data Availability}
The data that support the findings of this article are openly available~\cite{DataAvail}.

\appendix

\begin{figure}[t]
	\includegraphics[width=8.6cm]{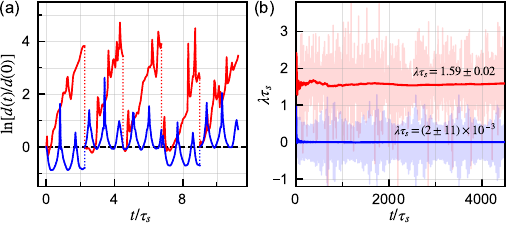}
 \caption{Numerical extraction of the LLEs for the representative regular and chaotic initial conditions, $\boldsymbol{x}_R$ (blue) and $\boldsymbol{x}_C$ (red), introduced in Sec.~III~A. (a) Time evolution of the logarithmic phase-space distance between two initially nearby trajectories, which is used to compute the LLE. (b) Cumulative estimate of the Lyapunov exponent (solid line) as a function of simulated time, together with the iteration-wise values (light line). The quoted uncertainties are the standard error of the mean value over the last $N_{\mathrm{iter}}/2$ iterations.}
	\label{FIG_A}
\end{figure}

\section{Single-trajectory analysis}

We determined the LLE using the standard rescaling method based on two nearby trajectories~\cite{Benettin80a,Skokos10}. Briefly, we consider a trajectory $\boldsymbol{x}(t)$ starting from $\boldsymbol{x}_0$ at $t=0$, as well as an adjunct trajectory $\boldsymbol{x}'(t)$ starting from a nearby point $\boldsymbol{x}'_0$ that is separated by $d_0$ from $\boldsymbol{x}_0$ in random orientation. We evolve both trajectories for a reset time $T_r$, and their separation $d_1$ gives the growth rate of separation as $\lambda_1=\frac{1}{T_r}\ln (d_1/d_0)$. Then, after adjusting $\boldsymbol{x}'$ to rescale the separation vector to magnitude $d_0$ while preserving its direction, we evolve the trajectories for another $T_r$ and estimate $\lambda_2$ from their resulting separation $d_2$. This process is repeated for $N_{\mathrm{iter}}$ iterations; the LLE for the trajectory $\boldsymbol{x}(t)$ is determined as $\lambda = \frac{1}{N_{\mathrm{iter}}/2}\sum_{n=N_{iter}/2+1}^{N_{\mathrm{iter}}}\lambda_n$ to reduce initial transient effects, and its numerical uncertainty is estimated as the standard error of the mean value.

The numerical parameters $d_0$, $T_r$, and $N_{\mathrm{iter}}$ were chosen to ensure that the extracted LLE is both physically meaningful and numerically stable. The initial separation $d_0$ must be small enough to probe the local instability of nearby trajectories, $d(t)\sim d_0 \, e^{\lambda t}$, yet not so small that numerical round-off or integration tolerances dominate. The reset time $T_r$ must be long enough for the separation to grow appreciably within each interval, while remaining well below the saturation scale set by the bounded accessible manifold. The number of reset iterations $N_{\mathrm{iter}}$ must then be large enough for the cumulative estimate to converge to a stable value with a reliable uncertainty, while retaining a moderate computational cost.
When determining the LLE, we used $d_0=10^{-6},\, T_r=0.05~\mathrm{s}(=2.25 \tau_s)$, and $N_{\mathrm{iter}}=2000$.
The total averaging time is $N_{\mathrm{iter}}T_r=100~\mathrm{s}=4500\,\tau_s$, which is sufficiently long to obtain a stable estimate of the LLE.

Figure~\ref{FIG_A}(a) shows the time evolution of the logarithmic distance between two initially nearby trajectories. For chaotic motion, it exhibits repeated exponential growth within each reset interval; for regular motion, the distance remains bounded and oscillatory. Figure~\ref{FIG_A}(b) shows the cumulative average of the iteration-wise exponents. The cumulative value converges to a positive constant for a chaotic trajectory and to zero for a regular one, while the residual fluctuations provide a practical estimate of the numerical uncertainty.

\begin{figure}[t]
    \includegraphics[width=8.6cm]{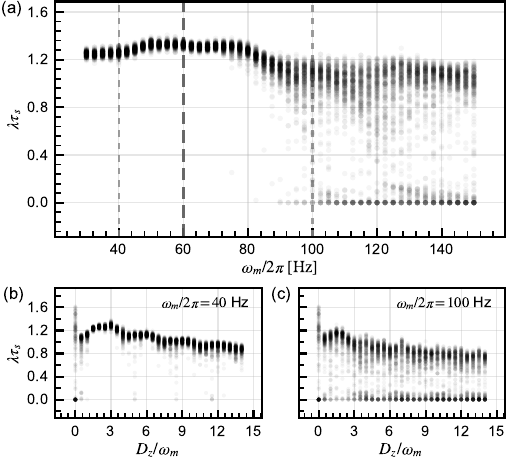}
    \caption{
    Modulation frequency dependence of the LLE distribution. (a) LLE values for 200 Haar-sampled initial conditions as a function of the modulation frequency \(\omega_m\) at fixed modulation amplitude \(\hbar D_z/\varepsilon_s = 2.2\). The dark dashed line marks the choice \(\omega_m/2\pi = 60~\mathrm{Hz}\) in the main text, and the light dashed lines indicate the representative low- and high-frequency cases shown in the following panels. (b)--(c) LLE distributions as functions of \(D_z/\omega_m\) at fixed \(\omega_m/2\pi = 40~\mathrm{Hz}\) and \(100~\mathrm{Hz}\), respectively. 
    }
    \label{Fig_B}
\end{figure}

The phase-space coverage of a trajectory was quantified by estimating the Shannon entropy from its coarse-grained occupation histogram, as described in Sec.~III A~\cite{Cincotta20}. 
Throughout this work, trajectories are sampled over $t=100~\mathrm{s}$ using the stored numerical solution at uniform intervals \(T_{\textrm{samp}}=1~\mathrm{ms}\), yielding \(N_s=100{,}000\).
We constructed a four-dimensional histogram of the sampled spin states over $(\rho_0,m,\theta_+, \theta_-)$ using uniform rectangular bins within the ranges $\rho_0\in[0,1]$, $m\in[-1,1]$, $\theta_+\in[-\pi,\pi)$, and $\theta_-\in[-\pi,\pi)$.
We used $8$ bins along each of the four histogram directions, corresponding to a total of $N_{\mathrm{bin}}=8^4=4096$ cells in the rectangular bounding grid.
The parameter values were chosen to balance two competing requirements: finer partitions improve spatial resolution but increase sampling noise, whereas coarser partitions suppress noise at the cost of washing out phase-space structure.

To determine the entropy deficit $\Delta S \equiv S_{\mathrm{Haar}} - S$, we calculated $S_{\mathrm{Haar}}$ from $N_s$ samples drawn from Haar-random pure spin-1 states, generated by sampling the three complex amplitudes from independent complex Gaussian variables and normalizing the resulting state vector~\cite{Zyczkowski01}.
For sufficiently large $N_s$, $S_{\mathrm{Haar}}\sim\log N_{\mathrm{bin}}$. Subtracting $S_{\mathrm{Haar}}$ therefore removes the leading-order dependence of $S$ on $N_{\mathrm{bin}}$, making the entropy deficit a more robust and physically transparent indicator of phase-space coverage. 
In a rough coarse-grained sense, $\mathcal{V} = \exp(-\Delta S)$ estimates the fraction of phase space occupied by the trajectory relative to the Haar reference.

\begin{figure}[t]
	\includegraphics[width=8.6cm]{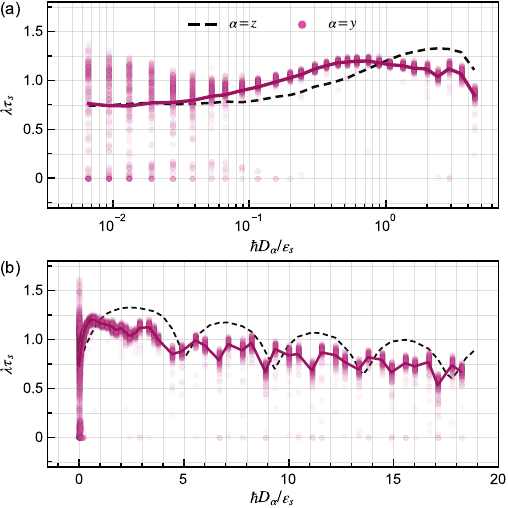}
	\caption{Modulation direction dependence of the LLE at $\omega_m/2\pi = 60~\mathrm{Hz}$. Faint magenta circles show the LLE values for 200 Haar-sampled initial conditions under $y$-modulation. The solid magenta and dashed black curves represent the mean LLE for $y$- and $z$-modulation, respectively. Panels (a) and (b) display the same data on logarithmic and linear horizontal scales, respectively, with panel (a) emphasizing the weak-modulation regime.}
	\label{Fig_C}
\end{figure}
\section{Modulation frequency dependence}

We investigated the dependence of the LLE distribution on the modulation frequency to complement the strong-driving analysis in Sec.~III~D. In Fig.~\hyperref[Fig_B]{9(a)}, we fix the modulation amplitude to $\hbar D_z = 2.2\varepsilon_s$ and evaluate the LLE for 200 Haar-sampled initial conditions while varying $\omega_m$. Around $\omega_m/2\pi = 60~\mathrm{Hz}$, the distribution remains relatively narrow, indicating that the system still forms an almost homogeneous chaotic sea. However, as the modulation frequency increases beyond approximately $90~\mathrm{Hz}$, the distribution broadens substantially and a significant fraction of the trajectories acquire near-zero LLE values. This indicates that the drive becomes less effective in sustaining global mixing at higher frequencies.

Figs.~\hyperref[Fig_B]{9(b)} and \hyperref[Fig_B]{9(c)} clarify this trend by showing fixed-frequency scans as functions of $D_z/\omega_m$ for $\omega_m/2\pi = 40~\mathrm{Hz}$ and $100~\mathrm{Hz}$, respectively. At $40~\mathrm{Hz}$, the LLE values remain relatively concentrated and the dip structure is only weakly resolved. At $100~\mathrm{Hz}$, by contrast, the distribution is much broader and extends down to near-zero LLE over a wide range of $D_z/\omega_m$.
This behavior is consistent with the observations from the rotating-frame picture presented in the main text. When $\omega_m$ is small, not only the leading $J_1$ channel but also higher-harmonic channels remain slow enough to mediate phase-space transport, so suppressing a single channel does not immediately destroy global mixing. By contrast, when $\omega_m$ is too large, even the lowest-order oscillating channels become too rapid to efficiently mediate transport across phase-space shells.

\section{Modulation direction dependence}

Motivated by the form of $\tilde{H}_F(t)$ in Eq.~(\ref{InteractionHamil}), we examined the mixing effect of transverse-field modulation.
Figure~\hyperref[Fig_C]{10} compares the LLE values for $y$- and $z$-directional modulations, where $H_F(t)=\hbar D_\alpha \sin (\omega_m t) f_{\alpha}$ with $\alpha=y,z$.
In the weak-modulation regime ($\hbar D_\alpha \lesssim \varepsilon_s$), $y$-modulation enhances the LLE more rapidly than $z$-modulation, indicating that transverse driving more efficiently destabilizes regular islands. However, $z$-modulation attains a higher maximum LLE. 
At larger amplitudes, both cases exhibit an overall decreasing and oscillatory trend. The sharp dips associated with Bessel zeros, prominent in $z$-modulation, are largely absent for $y$-modulation. 
Modulation along $x$ (not shown) was found to yield behavior similar to that in the $y$-modulation case.

The direction dependence originates from the quadratic Zeeman anisotropy, $q\boldsymbol{\zeta}^\dagger F_z^2\boldsymbol{\zeta}$. For $z$-axis modulation, $[\text{F}_z,\text{F}_z^2]=0$, and the spin-interaction term is rotationally invariant. Thus, in the comoving frame that removes the longitudinal modulation, the time dependence resides entirely in the transverse-field term, which is decomposed into harmonics
proportional to $\hbar\Omega J_n(\tilde D_z)$ [Eq.~(\ref{InteractionHamil})]. A zero of
$J_1(\tilde D_z)$ can thus eliminate the leading oscillating channel and
strongly suppress global phase-space transport.

For $y$-axis modulation, the comoving frame is defined by $U_y(t)=\exp\!\left[i\tilde D_y\cos(\omega_m t)\text{F}_y\right]$ with $\tilde D_y=D_y/\omega_m$. The transformed Hamiltonian is
\begin{widetext}
\begin{align*}
H_I^{(y)}(t)
&= \frac{\varepsilon_s}{2} |\boldsymbol{f}|^2  +\frac{q}{2}\boldsymbol{\zeta}^{\dagger} (\text{F}_z^2+\text{F}_x^2) \boldsymbol{\zeta} +\frac{q}{2}J_0(2\tilde D_y)\boldsymbol{\zeta}^{\dagger} (\text{F}_z^2-\text{F}_x^2)\boldsymbol{\zeta}  \\
& +q\sum_{n=1}^{\infty}(-1)^nJ_{2n}(2\tilde D_y)\cos(2n\cdot\omega_m t)\boldsymbol{\zeta}^{\dagger} (\text{F}_z^2-\text{F}_x^2)\boldsymbol{\zeta} \\
& -q\sum_{n=1}^{\infty}(-1)^nJ_{2n-1}(2\tilde D_y)\cos[(2n-1)\cdot\omega_m t]\boldsymbol{\zeta}^{\dagger} (\text{F}_z\text{F}_x+\text{F}_x\text{F}_z)\boldsymbol{\zeta}   \\
& +\hbar\Omega J_0(\tilde D_y)f_x +2\hbar\Omega\sum_{n=1}^{\infty}(-1)^nJ_{2n}(\tilde D_y)\cos(2n\cdot\omega_m t)f_x 
+2\hbar\Omega\sum_{n=1}^{\infty}(-1)^nJ_{2n-1}(\tilde D_y)\cos[(2n-1)\cdot\omega_m t]f_z.
\label{InteractionHamil_y}
\end{align*}
\end{widetext}
In this case, $[\text{F}_y,\text{F}_z^2]\neq0$, 
so the quadratic Zeeman term also generates harmonics proportional to $qJ_n(2\tilde D_y)$.
Consequently, even when the first-harmonic contribution $\propto  J_1(\tilde D_y)$ vanishes, the term $\propto qJ_1(2\tilde D_y)$ generally remains finite. The leading harmonic is therefore not canceled as completely as for $z$-axis modulation, explaining the much weaker Bessel-zero suppression.

In experiment, modulation along an arbitrary direction can be realized by controlling both the transverse rf magnetic field $\mathbf{B}_{\rm rf}(t)$ and bias field $B_z(t)$. Specifically, we consider $\mathbf{B}_{\rm rf}(t) = B_x(t)[\cos(\omega_L t)\,\hat{\mathbf{x}}+\sin(\omega_L t)\,\hat{\mathbf{y}}] + B_y(t)[\cos(\omega_L t)\,\hat{\mathbf{y}}-\sin(\omega_L t)\,\hat{\mathbf{x}}],$
where $\omega_L=\gamma B_0$ is the Larmor frequency with $\gamma=g_F\mu_B/\hbar$, $g_F$ being the Land\'e $g$-factor and $\mu_B$ the Bohr magneton. 
Within the rotating wave approximation, the Hamiltonian is given by 
$H= q\,\boldsymbol{\zeta}^\dagger \text{F}_z^2\boldsymbol{\zeta}+\frac{\varepsilon_s}{2}|\boldsymbol{f}|^2 + \hbar \boldsymbol{D}(t)\cdot \boldsymbol{f}$,
where $\boldsymbol{D}(t)=\gamma \big(B_x(t), B_y(t), \delta B_z(t) \big)$.


\begin{thebibliography}{99}

\bibitem{DAlessio16} L. D'Alessio, Y. Kafri, A. Polkovnikov, and M. Rigol, From quantum chaos and eigenstate thermalization to statistical mechanics and thermodynamics, Adv. Phys. \textbf{65}, 239 (2016).

\bibitem{Iyoda18} E. Iyoda and T. Sagawa, Scrambling of quantum information in quantum many-body systems, Phys. Rev. A \textbf{97}, 042330 (2018).

\bibitem{Brandao16} F. G. S. L. Brand\~ao, A. W. Harrow, and M. Horodecki, Local random quantum circuits are approximate polynomial-designs, Commun. Math. Phys. \textbf{346}, 397 (2016).

\bibitem{PilatowskyCameo24} S. Pilatowsky-Cameo, I. Marvian, S. Choi, and W. W. Ho, Hilbert-Space Ergodicity in Driven Quantum Systems: Obstructions and Designs, Phys. Rev. X \textbf{14}, 041059 (2024).

\bibitem{Choi23} J. Choi, A. L. Shaw, I. S. Madjarov, X. Xie, R. Finkelstein, J. P. Covey, J. S. Cotler, D. K. Mark, H.-Y. Huang, A. Kale, H. Pichler, F. G. S. Brand\~ao, S. Choi, and M. Endres, Preparing random states and benchmarking with many-body quantum chaos, Nature \textbf{613}, 468 (2023).

\bibitem{Eckmann85} J.-P. Eckmann and D. Ruelle, Ergodic theory of chaos and strange attractors, Rev. Mod. Phys. \textbf{57}, 617 (1985).

\bibitem{Meiss92} J.~D. Meiss, Symplectic maps, variational principles, and transport, Rev. Mod. Phys. \textbf{64}, 795 (1992).

\bibitem{Rautenberg20} M. Rautenberg and M. G\"arttner, Classical and quantum chaos in a three-mode bosonic system, Phys. Rev. A \textbf{101}, 053604 (2020).

\bibitem{Tabor89} M. Tabor, \textit{Chaos and Integrability in Nonlinear Dynamics: An Introduction} (Wiley, New York, 1989).

\bibitem{Kim24} J. Kim, J. Jung, J. Lee, D. Hong, and Y. Shin, Chaos-assisted turbulence in spinor Bose-Einstein condensates, Phys. Rev. Research \textbf{6}, L032030 (2024).

\bibitem{Jung23} J. H. Jung, J. Lee, J. Kim, and Y. Shin, Random spin textures in turbulent spinor Bose-Einstein condensates, Phys. Rev. A \textbf{108}, 043309 (2023).

\bibitem{Hong23} D. Hong, J. Lee, J. Kim, J. H. Jung, K. Lee, S. Kang, and Y. Shin, Spin-driven stationary turbulence in spinor Bose-Einstein condensates, Phys. Rev. A \textbf{108}, 013318 (2023).

\bibitem{Benettin80a} G. Benettin, L. Galgani, A. Giorgilli, and J.-M. Strelcyn, Lyapunov characteristic exponents for smooth dynamical systems and for Hamiltonian systems; A method for computing all of them. Part 1: Theory, Meccanica \textbf{15}, 9 (1980).

\bibitem{Skokos10} C. Skokos, The Lyapunov characteristic exponents and their computation, Lect. Notes Phys. \textbf{790}, 63 (2010).

\bibitem{Cincotta20} P.~M. Cincotta and I.~I. Shevchenko, Correlations in area preserving maps: A Shannon entropy approach, Physica D \textbf{402}, 132235 (2020).

\bibitem{Ho22} W. W. Ho and S. Choi, Exact emergent quantum state designs from quantum chaotic dynamics, Phys. Rev. Lett. \textbf{128}, 060601 (2022).

\bibitem{Cotler23} J. S. C. Cotler, D. K. Mark, H.-Y. Huang, F. Hern\'andez, J. Choi, A. L. Shaw, M. Endres, and S. Choi, Emergent quantum state designs from individual many-body wave functions, PRX Quantum \textbf{4}, 010311 (2023).


\bibitem{Bukov15} M. Bukov, L. D'Alessio, and A. Polkovnikov, Universal High-Frequency Behavior of Periodically Driven Systems: from Dynamical Stabilization to Floquet Engineering, Adv. Phys. \textbf{64}, 139 (2015).

\bibitem{Eckardt17} A. Eckardt, Colloquium: Atomic quantum gases in periodically driven optical lattices, Rev. Mod. Phys. \textbf{89}, 011004 (2017).

\bibitem{Mark24} D. K. Mark, F. Surace, A. Elben, A. L. Shaw, J. Choi, G. Refael, M. Endres, and S. Choi, Maximum Entropy Principle in Deep Thermalization and in Hilbert-Space Ergodicity, Phys. Rev. X \textbf{14}, 041051 (2024).

\bibitem{Ho98} T.-L. Ho, Spinor Bose Condensates in Optical Traps, Phys. Rev. Lett. \textbf{81}, 742 (1998).

\bibitem{Kawaguchi12} Y. Kawaguchi and M. Ueda, Spinor Bose-Einstein condensates, Phys. Rep. \textbf{520}, 253 (2012).

\bibitem{StamperKurn13} D. M. Stamper-Kurn and M. Ueda, Spinor Bose gases: Symmetries, magnetism, and quantum dynamics, Rev. Mod. Phys. \textbf{85}, 1191 (2013).

\bibitem{Law98} C. K. Law, H. Pu, and N. P. Bigelow, Quantum Spins Mixing in Spinor Bose-Einstein Condensates, Phys. Rev. Lett. \textbf{81}, 5257 (1998).

\bibitem{Yi02} S. Yi, \"O. E. M\"ustecaplio\u{g}lu, C. P. Sun, and L. You, Single-mode approximation in a spinor-1 atomic condensate, Phys. Rev. A \textbf{66}, 011601(R) (2002).

\bibitem{Bengtsson17} I. Bengtsson and K. Zyczkowski, Geometry of Quantum States: An Introduction to Quantum Entanglement, 2nd ed. (Cambridge University Press, Cambridge, 2017).

\bibitem{Zyczkowski01} K. \.{Z}yczkowski and H.-J. Sommers, Induced measures in the space of mixed quantum states, J. Phys. A: Math. Gen. \textbf{34}, 7111 (2001).

\bibitem{Yi03} S. Yi, \"O. E. M\"ustecapl{\i}o\u{g}lu, and L. You, Quantum Phase Diffusions of a Spinor Condensate, Phys. Rev. Lett. \textbf{90}, 140404 (2003).

\bibitem{Leslie09} S. R. Leslie, J. Guzman, M. Vengalattore, J. D. Sau, M. L. Cohen, and D. M. Stamper-Kurn, Amplification of fluctuations in a spinor Bose-Einstein condensate, Phys. Rev. A \textbf{79}, 043631 (2009).

\bibitem{Altmann06} E. G. Altmann, A. E. Motter, and H. Kantz, Stickiness in Hamiltonian systems: From sharply divided to hierarchical phase space, Phys. Rev. E \textbf{73}, 026207 (2006).

\bibitem{Zou07} Y. Zou, M. Thiel, M. C. Romano, and J. Kurths, Characterization of stickiness by means of recurrence, Chaos \textbf{17}, 043101 (2007).

\bibitem{Lozej20} \v{C}. Lozej, Stickiness in generic low-dimensional Hamiltonian systems: A recurrence-time statistics approach, Phys. Rev. E \textbf{101}, 052204 (2020).

\bibitem{Altmann13} E.~G.~Altmann, J.~S.~E.~Portela, and T.~T\'el, Leaking chaotic systems, Rev. Mod. Phys. \textbf{85}, 869 (2013).

\bibitem{Brinkmann16} A. Brinkmann, Introduction to average Hamiltonian theory. I. Basics, Concepts Magn. Reson. Part A \textbf{45A}, e21414 (2016).

\bibitem{Rahav03} S. Rahav, I. Gilary, and S. Fishman, Effective Hamiltonians for periodically driven systems, Phys. Rev. A \textbf{68}, 013820 (2003).

\bibitem{Dunlap86} D.~H. Dunlap and V.~M. Kenkre, Dynamic localization of a charged particle moving under the influence of an electric field, Phys. Rev. B \textbf{34}, 3625 (1986).

\bibitem{Grifoni98} M. Grifoni and P. H\"anggi, Driven quantum tunneling, Phys. Rep. \textbf{304}, 229 (1998).

\bibitem{Lignier07} H. Lignier, C. Sias, D. Ciampini, Y. Singh, A. Zenesini, O. Morsch, and E. Arimondo, Dynamical Control of Matter-Wave Tunneling in Periodic Potentials, Phys. Rev. Lett. \textbf{99}, 220403 (2007).

\bibitem{DataAvail} J. Kim, M. Jeong, J. Han and Y. Shin, Data and analysis scripts for "Efficient ensemble randomization by tuning chaos in a nonlinear spin-1 system" [Data set], Zenodo (2026), doi:10.5281/zenodo.24161774.

\end{thebibliography}
\end{document}